\documentclass[useAMS,usenatbib]{mn2e}
\usepackage{graphicx}
\usepackage{subfigure}
\usepackage{latexsym}
\usepackage{natbib}
\usepackage{txfonts}
\bibpunct{(}{)}{;}{a}{}{,}

\newcommand{\apj}{ApJ}

\newcommand{\apjl}{ApJ}
\newcommand{\mnras}{MNRAS}
\newcommand{\pasj}{PASJ}
\newcommand{\aap}{A\&A}
\newcommand{\nat}{Nature}

\newcommand{\araa}{ARA\&A}
\newcommand{\apss}{Ap\&SS}

\newcommand{\text}[1]{\quad\mbox{#1}\quad}

\newcommand{\sub}[1]{_{\mbox{\tiny #1}}}

\newcommand{\mean}[1]{\langle{#1}\rangle}

\begin{document}

\def\ls{LS~5039}
\def\lsi{LS~I~$+$63$-$303}
\def\psr{PSR~B1259$-$63/LS2883}
\def\cyg{Cygnus~X-1}
\def\muq{$\mu$Q}
\title{Direct Wind Accretion and Jet Launch in Binary Systems}

\author[Barkov et al]{Maxim V.~Barkov${}^{1,2}$,
     Dmitry V.~Khangulyan${}^{3}$\\
${}^1${Max-Planck-Institut f\"ur Kernphysik, Saupfercheckweg 1, 69117 Heidelberg, Germany}\\
${}^2$ Space Research Institute, 84/32 Profsoyuznaya Street, Moscow
117997, Russia\\
${}^3$ Institute of Space and Astronautical Science/JAXA, 3-1-1 Yoshinodai, Chuo-ku, Sagamihara, Kanagawa 252-5210, JAPAN}

\date{}
\maketitle
\begin{abstract}
  In this paper we study the wind accretion onto a rotating
  black hole in the close binary system harboring a young massive star. It
  is shown that the angular momentum of the accreted stellar
  wind material is not sufficient for the formation of an accretion
  disk. On the other hand, in the considered conditions the
  Blanford-Znajek mechanism can be activated, thus powerful jets can
  be launched in the direction of the rotation axis of the black hole. 
  Importantly, no observational signatures of accretion, as
  typically seen from the thermal X-ray emission from the accretion
  disks, are expected in the suggested scenario. Here, properties of the
  generated jet are studied numerically in the framework of a
  2D general relativity magnetohydrodynamical approach. Due to the
  accumulation of the magnetic flux at the black hole horizon, the jet
  power is expected to be modulated on a sub-second
  time-scale. Although the intervals between jet active phases  depend 
  on the magnetic flux escape from the
  black hole horizon (which can be modeled self-consistently only using a
  3D code), a general estimate of the averaged jet power is
  obtained.  It is expected that for the black hole rotation, expected
  in stellar binary system (the dimensionless rotation parameter
  $a=0.5$), approximately $10\%$ of the accreted rest energy can be
  channeled into the jets. In the specific case of the gamma-ray binary
  system \ls, the obtained jet luminosity can be responsible for the
  observed GeV radiation if one invokes Doppler boosting,
  which can enhanced the apparent flux from the system.
\end{abstract}


\section{Introduction}
\label{intro}
Several binary systems, consisting of a massive star and a compact
object (CO) --neutron star (NS) or black hole (BH)-- were detected in
the very high energy (VHE) regime
\citep{hess05,hess05a,hess06,hess09,magic06,magic07,magic09,veritas08,veritas09,veritas11},
and their non-thermal emission is believed to be linked to the
processes related to the CO \citep[see e.g.][for a
review]{bk09}. Moreover, given large values of gamma-gamma opacity
expected in the close vicinity of CO, the VHE gamma-ray production
most likely should be related in the outflow originated in the CO. In
particular, in the system \psr{} a pulsar is orbiting a massive O-type
star \citep[see][and reference therein]{nrh11}, and the detected X-ray
and TeV radiation is attributed to synchrotron and inverse Compton
(IC) emission produced by electrons accelerated at the pulsar wind
termination shock \citep{kbs00,kha07,utt09}, while emission detected
in the GeV energy range \citep{fermi_psr11,fermi_psr11a} can be
explained by the bulk Comptonizaiton of the pulsar wind
\citep{kab11a}. In the case of \cyg, the evidence of detection in TeV found
MAGIC \citep{magic07} is most likely related to the non-thermal activity occurring in
the jet launched by a black hole \citep{enc98} accreting matter from a
O9.7Iab star \citep{zio05}.

The classification of compact objects (CO) in these two cases is
rather clear since (i) pulsed radio emission coherent with the orbital
motion is detected from \psr{} \citep{jml96}; and (ii) in the case of
\cyg{} the mass of the CO exceeds the Oppenheimer-Volkoff upper limit
for the mass of NS \citep{cas07}, and the accretion disk is clearly
seen in the X-ray energy band \citep{enc98}.

The physical processes responsible for the VHE radiation detected
from two other systems, \ls{} and \lsi, is still debated, since
currently there is no observational evidences which allows robust
conclusions regarding the nature of the CO. In particular, despite
intensive searches no pulsed emission has been detected from these
systems. Although, strong free-free absorption, expected in these
relatively compact binaries, could hide the radio pulsations
\citep{dub06}, the strict upper limits obtained in the X-ray energy
band \citep{rtk10,rtc11} pose additional constraints for the
realization of the binary pulsar scenario in these systems. Another,
independent on the orientation of the ``pulsar beam'', another test for the
presence of a pulsar in the binary system is related to the impact of
the pulsar wind on the stellar environment. Namely, the collision of
the stellar and relativistic pulsar winds should give rise not only to
the non-thermal emission related to the pulsar wind, but the stellar
wind would be also shocked producing thermal X-rays.  
This approach was used to constrain the spin-down
(SD) luminosity of the possible pulsar in the case of \ls{}
\citep{zbp11}. Importantly, the derived upper limits appeared to be
very close to the non-thermal luminosity detected with {\it Fermi} Large Area
Telescope ({\it Fermi}/LAT) \citep{fermi09a}. Thus, one should expect
either a perfect conversion of the pulsar energy to GeV gamma-ray, or
to invoke Doppler boosting of the emission. Although, the
relativistic motion of the shocked pulsar wind is a rather natural
effect in binary pulsar systems \citep{bkk08,kab08,dch10}, fast mixing
of the stellar and pulsar winds may destroy the relativistic regime of
the outflow \citep{brb11}, and the efficient Doppler boosting of the
emission is possible only on the scales limited by the binary system
size.

Orbital phase dependent extended radio structures, predicted
for the binary pulsar systems \cite{dub06}, were found in the
gamma-ray binary systems \psr, \ls{} and \lsi{} 
\citep{dmr06,rpm08,mjr11,mrp11}. This may be
considered as a strong evidence for the presence of pulsars in these
systems, however (i) the extended radio emission can be naturally
produced in gamma-ray binary systems via the synchrotron emission of
secondary pairs \citep{bka08,bk11}; (ii) to interpret the resolved 
structures requires a knowledge of the flow dynamics that presently lacks. 
This uncertainty can be to some extend
explained by the excessive simplification adopted in the theoretical
calculations, in particular the interaction of the winds in binary
pulsar can be rather complicated \citep{bkk08,brb11} and differ
significantly \citep{bkk11} from the case realized at the interaction
of a pulsar with the interstellar medium \citep{kc84a,dub06}. Thus, the
interpretation of these observations in the context of the pulsar 
scenario remains inconclusive.

An alternative scenario implies the production of VHE gamma-rays in
miroquasar jets launched by an accreting black hole.  Although 
both systems, \ls{} and \lsi, do not show thermal emission produced by
the accreation disk, as it is expected in the classical
microquasar (\muq) paradigm, the spectroscopic observations of \ls{}
favor a mass of the CO \citep{crr05,ssk11} exceeding the
Oppenheimer-Volkoff upper limit for the mass of NS. If a BH is indeed
located in these systems, the absence of the thermal emission of the
accretion disk might be interpreted as a signature of the {\it
radiatively non-efficient} disk accretion \citep{ny94,bb99}, although
the applicability of these scenarios is debated for sub-Eddington
stellar systems \citep{abin06}.  Therefore, a detailed study of the
properties of the accretion flow in binary system is required for
meaningful conclusions regarding the possibility to power these
sources, \ls{} and \lsi, by accretion. We show below that in these
systems of massive stars, which do not fill the Roshe lobe and the CO is 
not very massive, the accretion onto the BH should occur from the stellar wind.

For the sake of specificity, we adopt in this work the system 
parameters similar to those of \ls. 
In this system the CO is orbiting an O6.5V star in a
relatively non-eccentric orbit (see the system parameters in
Tab.~1). This system was detected as a bright non-thermal source in a
very broad energy range, from radio to VHE \citep[see e.g.][and
references therein]{bk09}. The spectral energy distribution (SED) of
the radiation is dominated by GeV gamma-rays. Namely, a flux of about
few$\times10^{35}\rm \, erg\,cm^{-2}s^{-1}$ was detected with {\it Fermi}/LAT
\citep{fermi09}. The fluxes in X-ray and TeV energy bands are
comparable and at the level of $\sim10^{34}\rm \,
erg\,cm^{-2}s^{-1}$ \citep{hess06,tku09}. X-ray, GeV and TeV emission
components were shown to be periodic and stable on year time-scales
\citep{hess06,ktu09,fermi09}. We note that the emission detected from
\lsi{} is known to differ significantly from orbit to
orbit \citep{ltz11,veritas11}.

According to our analytical estimates, the accretion rate can be relatively
high, in agreement with the results obtained by \citet{oor10}. 
Importantly, due to the lack of angular momentum,
the accretion proceeds without the formation of an accretion disk,
i.e. one should not expect detectable thermal X-rays from
the in-falling matter. Despite the absence of the accretion disk, our study 
shows that a powerful jet can be launched in this 
accretion regime. The crucial conditions for the formation of a powerful jet
are (i) magnetization of the in-falling flow; and (ii) high enough accretion
rate. If these conditions are fulfilled, a fast rotating BH can power
and launch a magnetically driven jet \citep{kb09}. Accretion of the
matter leads to the accumulation of the magnetic flux near the BH. When
the B-field reaches the critical value, the magnetic pressure pushes
out the in-falling matter pausing the accretion and dis-power the
jet. This phenomenon is similar to the well known {\it magnetically
  arrested disks} \citep{bkr74,bkr76,ktl92,nia03,I08,ruk09,
  tnm11}. The periodic stops of the central engine leads to a
formation of a quasi-periodic structure of the jet on a scale
significantly smaller (by three orders of magnitude) than the system
semi-major axis.

The paper is organized as following: in Sect.\ref{model} we present
the general properties of the accretion flow in binary systems; in
Sect.~\ref{res} we develop a 2D general relativity
magnetohydrodynamical (GRMHD) numerical model for the accretion flow,
and study different accretion regimes; in Sect.~\ref{concl} we discuss
the implications of the obtained solutions; and in Sect.~\ref{sec:sum}
we summarized the obtained results.

\section{Modeling accretion}
\label{model}
\subsection{General Properties of the Accretion Flow}
\label{gflow}
\label{mls5039}

In a binary system harboring the BH, and in which the optical component does
not fill the Roshe lobe, the BH will accrete from the stellar wind
directly. The accretion rate can be estimated from the Bondi-Hoyle
solution, which describes the accretion onto a center of mass $M$
moving with a constant speed $\varv_\infty$
%
%
through a uniform medium of density $\rho_\infty$.  The Bondi-Hoyle
mass accretion rate is well approximated by the equation
\begin{equation}
\dot{M}\sub{BH} = \pi R\sub{A}^2\rho_\infty\varv_\infty\,,
\label{bondi-hoyle}
\end{equation}
where
\begin{equation}
R\sub{A}=\frac{2GM}{\varv_\infty^2}
\label{ra1}
\end{equation}
is the accretion radius. In the Bondi-Hoyle solution the mean angular
momentum of accreted matter is zero.

The wind accretion can be only marginally described by the Bondi-Hoyle
solution due to the density and velocity gradients across the
BH trajectory, although provided that the
gradients are small on the scale of the accretion radius, the above
expression for $\dot{M}$ is still quite accurate \citep{im93,r97,r99}.
However, the accreted matter can obtain a non-zero mean angular
momentum \citep{is75,im93,dp80,r97,r99}. 
In analogy to the work by \cite{BK11rns} given the unsettled nature of this
issue, we will assume that the mean specific angular momentum of
accreted matter inside the accretion cylinder is
\begin{equation}
\mean{j\sub{A}}=\frac{\eta}{4}\Omega R\sub{A}^2,
\label{ja}
\end{equation}
where  $\Omega$ is the angular velocity of the orbital motion; and $\eta$ is a free parameter ($|\eta|<1$), which reflects our current ignorance.
\begin{table}
\caption{The parameters of the system  LS5039}  
 \begin{tabular}{   l    l    l }
\hline
\hline
Description & Designation & Value \\
 \hline
 Mass of star & $M_s$ & $26 M_{\odot}$\\
 Radius of star & $R_s$  & $9.3 R_{\odot}$ \\
 Temperature of the star & $T_s$ & 39, 000 K\\
 Stellar Wind termination velocity & $V_{\infty}$  & 2, 400 km/s \\
 Stellar Wind loss rate & $\dot{M}_s$  & $4\times10^{-7} \rm {M_{\odot} }{{yr}^{-1}}$ \\
 Orbital period & $P_s$  & 3.9 day \\
 Eccentricity of the orbit & $e$  & 0.24 \\
 The mass of the BH & $M_{BH}$  & $3 M_{\odot}$ \\
 Semimajor axis & $a_o$  & $3.5 R_s$ \\ 
\hline
 \end{tabular}
 \label{tabls}
 \end{table}

In the specific case of \ls, the above  estimates allow us to obtain analytically approximate values for
the accretion rate and angular momentum of the flow. The physical properties of the system are
summarized in Table.\ref{tabls} \citep[see][and the references therein]{crr05,ssk11}. 

The wind velocity profile can be approximated as \citep{kp00sw}
\begin{equation}\varv_{\rm w}(R)\approx\varv_{\infty} \left(1-\frac{R_s}{R}\right). 
\label{ws}
\end{equation}
Thus, the typical wind velocity at the orbital separation distance, $a_{\rm o}$, can be estimates as
$\varv_{\rm w}(a_o)\approx 1.7\times 10^8$ cm s$^{-1}$ (see Table.~1, for the used parameter values). This
value exceeds significantly the orbital velocity of the BH, $\varv_{\rm   o}\approx 4\times 10^{7}$ cm
s$^{-1}$, thus in the calculation one can safely neglect the BH velocity. Substituting the wind speed to
 Eq.(\ref{bondi-hoyle}) one obtains the value for the BH accretion rate: 
\begin{equation}
\dot{M}\sub{BH}=1.5\times 10^{-11}M_{\odot} \mbox{   yr}^{-1}\,.
\label{eq_m_dot}
\end{equation}
We note that this value depends on the star separation distance and  can vary up to a factor of
$3$-$5$ along the orbit \citep{oor10}. Note, that the accretion rate in 
the system \ls{}  is significantly smaller than in the case of Cygnus X$-$1
due to large difference in the masses of CO and in the speed of the stellar wind.  

Follow Eq.(\ref{ra1}), which gets $R_A\approx2.7\times10^{10}$~cm,
after it substitution in
Eq.(\ref{ja}) one can obtain the value for the angular momentum of the
accreted matter $j\sub{A}=3.3\times10^{15} \mbox{ cm}^2 \mbox{
  s}^{-1}$.  This value appears to be small as compared to the
specific angular momentum related to the last marginally bound orbit,
$j_{\rm mb}\approx 3.4 r_g c=3.4 GM_{BH}/c\approx 4.5\times10^{16}
\mbox{ cm}^2 \mbox{ s}^{-1}$ (here we assume the BH dimensionless
rotation parameter to be $a=0.5$) \citep{bpt72}. This relation,
$j\sub{A} \ll j_{\rm mb} $, implies that the accretion disk cannot be
formed in such an accretion regime, and one should expect a nearly
spherically symmetric in-flow of the stellar material. This allows to
perform the calculations of the accretion under 2D approximation.
Namely, we study the case of the spherical accretion onto a rotating
black hole \citep[the setup is similar to ones considered
by][]{kb09,BK10}. The initial magnetic field is assumed to be
homogeneous on the scale of the computational domain $R_{\rm
  com}=2\times10^9$~cm, and parallel to the rotation axis of BH. The
accretion rate was assumed to be steady on the computational
time-scale, and two values were adopted $\dot{M}\sub{BH}=(1.5{\rm \,\,
  and \,\,}8)\times 10^{-11}M_{\odot} \mbox{ yr}^{-1}\,.  $

\subsection{Numerical results}
\label{res}

The main details of our numerical method and various test simulations are described in the
literature \citep[see for details][]{K99,K04b,K06,KB07}. Here we outline just the key elements of
the approach.  The calculations are performed in  Kerr-Schild spacetime coordinates. The
computational grid was selected to be uniform in polar angle, $\theta$, where it has 320 cells and
logarithmic in spherical radius, $r$, where it has 773 cells.  The inner boundary is located just
inside the event horizon and adopts the free-flow boundary conditions. The outer boundary is located
at $r=2.2\times 10^9$~cm and at this boundary the flow is prescribed according to the free-fall
Bondi model with constant accretion rate.

In our study we check the feasibility the Blanford-Znajek (BZ) driven
jet formation \citep{rw75,lov76,BZ77} in the specific conditions of
wind accretion in the gamma-ray binary system
\ls. Simple relations obtained by \citet{BK08b} relate the BZ energy
realize to the magnetic flux
\begin{equation}
L\sub{BZ}=1.4\times10^{35} f(a) \Psi\sub{20}^2
\left(\frac{M_{BH}}{10 M_\odot}\right)^{-2} \,\mbox{erg}\, \mbox{s}^{-1},
\label{e-bz}
\end{equation}
where $f(a)=a^2\left(1+\sqrt{1-a^2}\right)^{-2}$ is a dimensionless
function accounting for the BH rotation; and $\Psi_{20}=\Psi/10^{20}
\mbox{ G cm}^2$ is the magnetic flux. According to \cite{BK10}, the
expected value of the dimensionless rotation parameter of the BH in a
stellar binary system should be relatively small, namely $a=0.5$
($f(a)=0.072$). 

Regarding the strength of the initial magnetic field, to reduce the calculation 
time we have assumed
a fixed value of $10^3$~G, which exceeds significantly the value of a
few G expected in stellar binary systems. We note that the actual
value of the magnetic field strength affects only the duration of the
magnetic flux accumulation stage, but does not influence the energy
release at the saturation phase (see details below). Thus, the
assumption of a strong magnetic field does not influence our results
while allowing to reduce significantly the computation time.

\begin{figure*}
\includegraphics[width=0.48\linewidth]{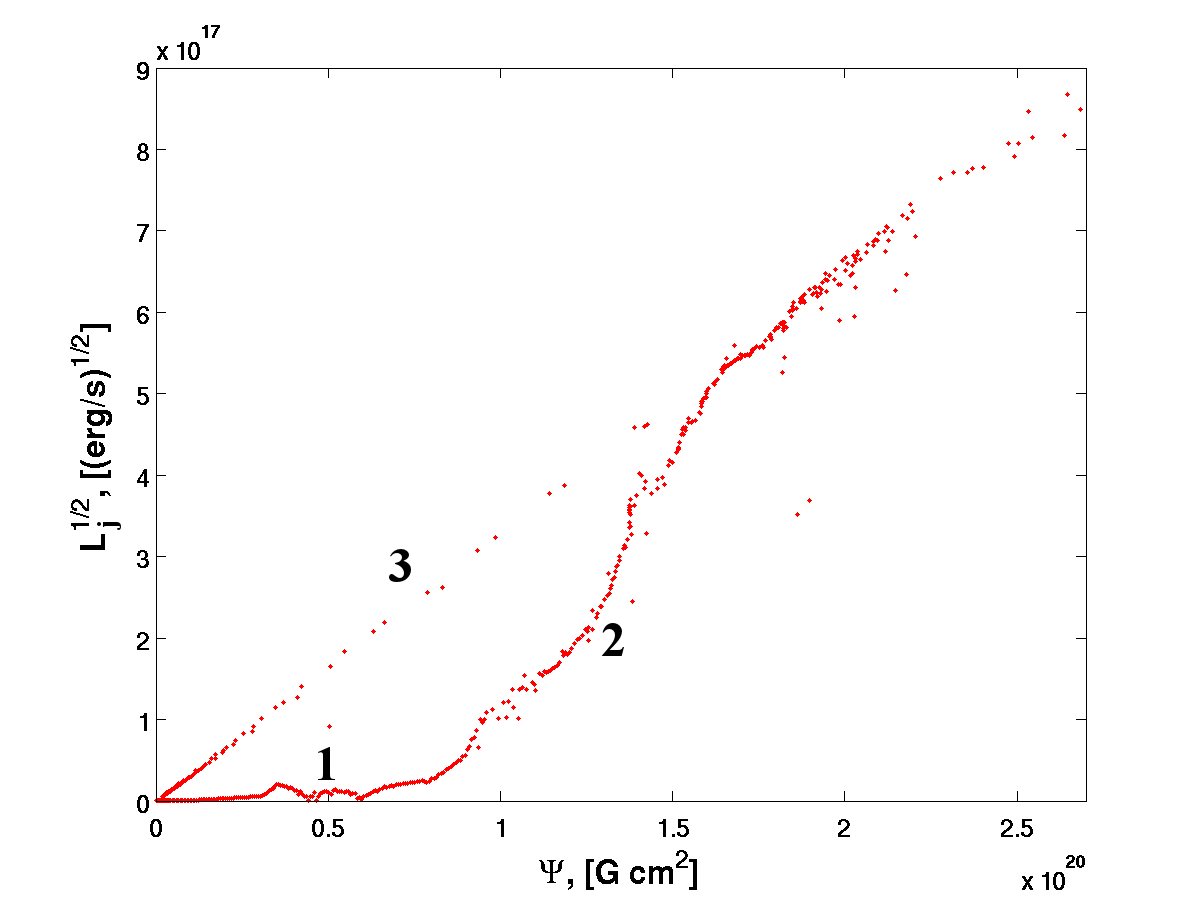}
\includegraphics[width=0.48\linewidth]{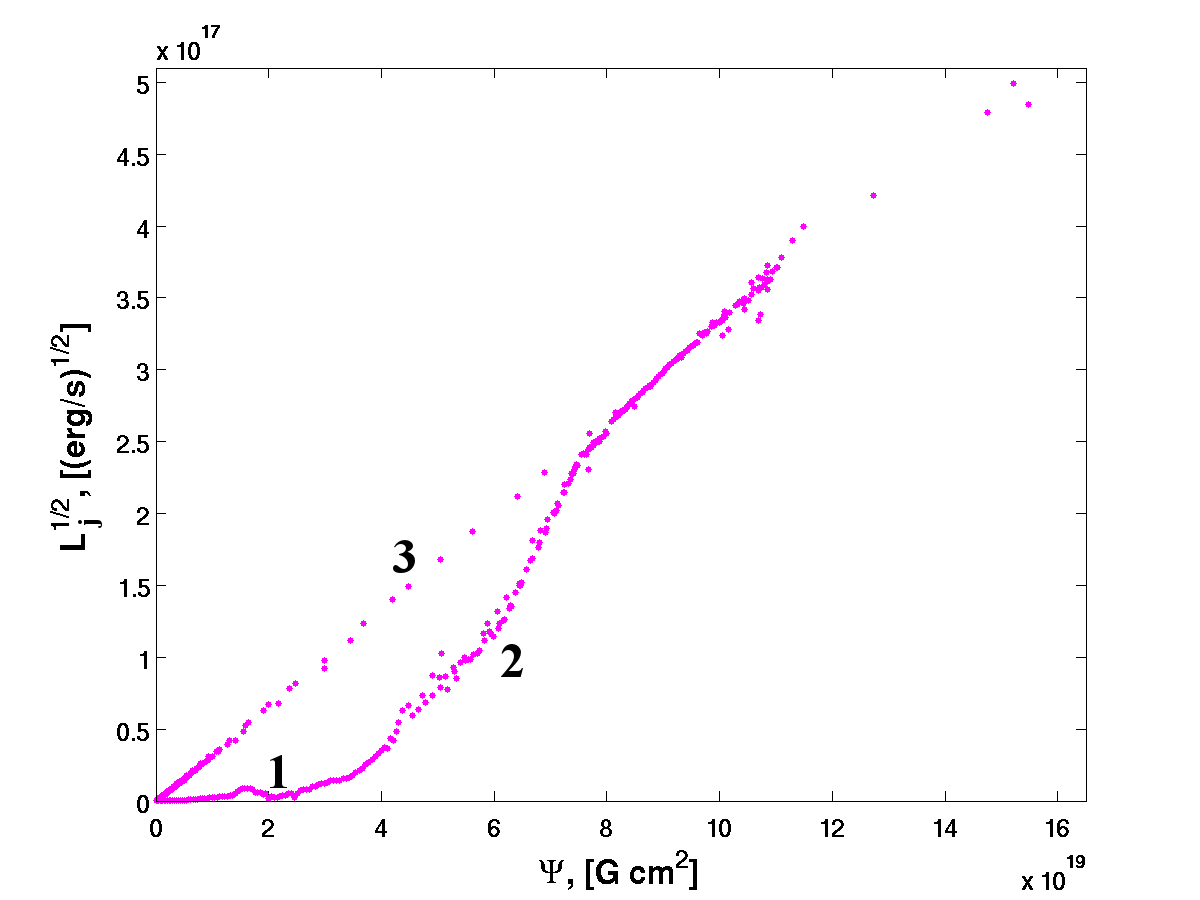}
\caption{ The square root of the jet luminosity as a function of the magnetic flux,
  $\Psi$, accumulated at the BH horizon. Three different regimes can
  be seen in the figure: (i) the initial accumulation of the magnetic
  flux (marked with ``1''); (ii) transition to the BZ regime (marked
  with ``2''); and (iii) the conventional BZ regime, when a linear
  dependence $L_{\rm j}^{1/2}\propto \Psi$ is expected (marked with
  ``3'').  The initial magnetic field in the stellar wind was assumed
  to be $10^3$~G, and two accretion rates were considered
  $\dot{M}_{BH}=8\times10^{-11} M_{\odot} \mbox{yr}^{-1}$ (left
  panel); and $\dot{M}_{BH}=1.5\times10^{-11} M_{\odot}
  \mbox{yr}^{-1}$ (right panel).  }
\label{lbzpsi}
\end{figure*}

At the initial stage of  the simulation  spherically symmetric accretion occurs and the magnetic
flux accumulates on the horizon of the BH. At the moment when the magnetic flux exceeds the critical
value \citep[see][for more details]{BK08c,kb09}, the BZ mechanism gets  activated and a magnetically
driven jet is launched. The accumulation phase can be clearly seen in Fig.~\ref{lbzpsi}, as the
lower trajectory in the {\it magnetic flux vs square root of the jet power} phase space. Namely,
during this period the system is characterized by an increasing magnetic flux without any ejection
of energy. This is  caused a too low value of the magnetic torque to activate the BZ
mechanism. 

The flux accumulation stage (marked with number 1 in Fig.\ref{lbzpsi}) continues until the magnetic
field flux reaches the critical value of $\Psi\approx 10^{20} \mbox{G cm}^2$ for an 
accretion rate of $\dot{M}_{BH}=8\times 10^{-11} M_{\odot} \mbox{yr}^{-1}$ (see transition in
Fig.\ref{lbzpsi}, left panel).  This value of magnetic flux can be used to check the feasibility of
the jet activation in binary systems with relatively weak magnetic field (we remind that in the
calculations we have assume a large value of B-field for the sake of reducing of the computational
expenses). The required value can be reached even in the case of accretion of an uniform magnetic field
of strength  $B_w\sim 0.1 \dot{M}^{1/2}_{-10}$~G from the radius $R_{\rm A}\simeq3\times10^{10}\,\rm
cm$. We note that stellar winds can carry significantly stronger magnetic fields
\citep[see eq.][]{um92}.

Once the jet is launched, the dense falling plasma is evacuated from the BH magnetosphere, and the BZ
mechanism starts to operate in the conventional way. Thus, the jet power should be described by
Eq.~(\ref{e-bz}). Indeed, in Fig.~\ref{lbzpsi} one can see that after the accumulation/cleaning
(marked with numbers 1 and 2 in the figure) stages the system remains on a straight line $L_{\rm
BZ}^{1/2}\propto \Psi$ (marked with number 3 in the figure).

\begin{figure*}
\includegraphics[width=0.48\linewidth]{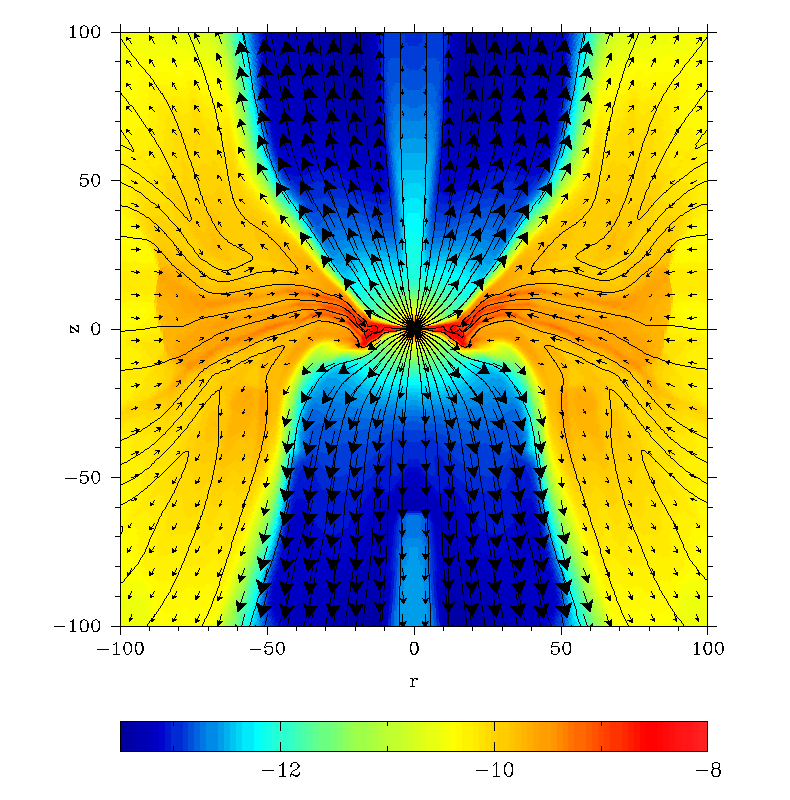}
\includegraphics[width=0.48\linewidth]{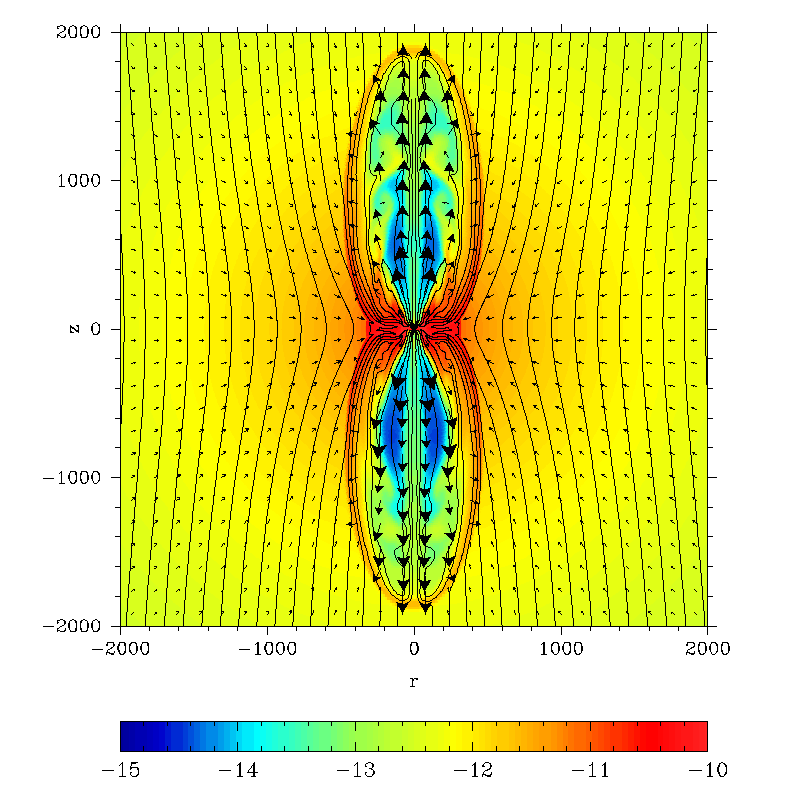}
\includegraphics[width=0.48\linewidth]{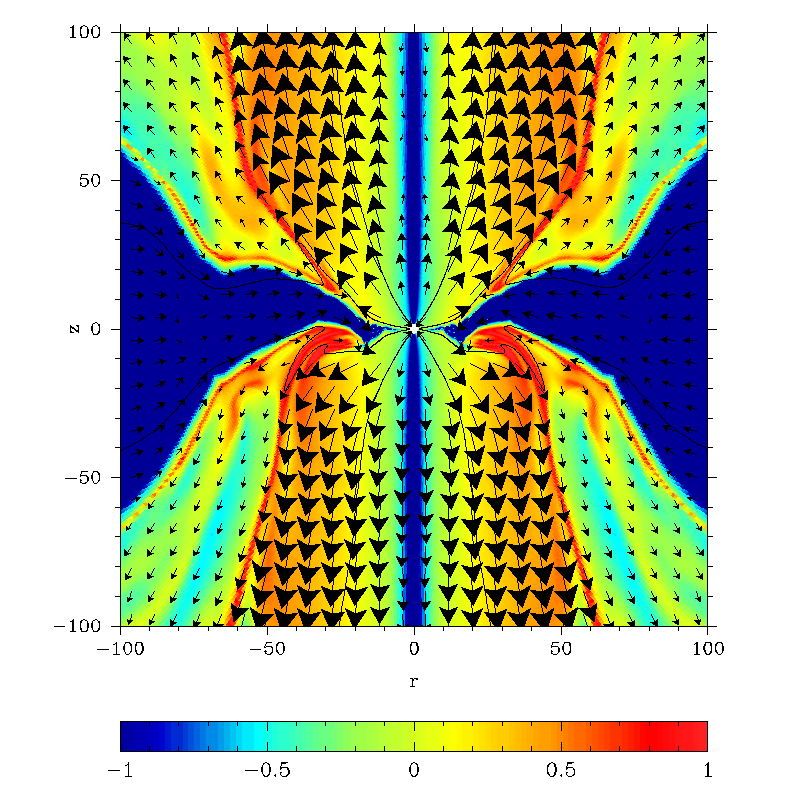}
\includegraphics[width=0.48\linewidth]{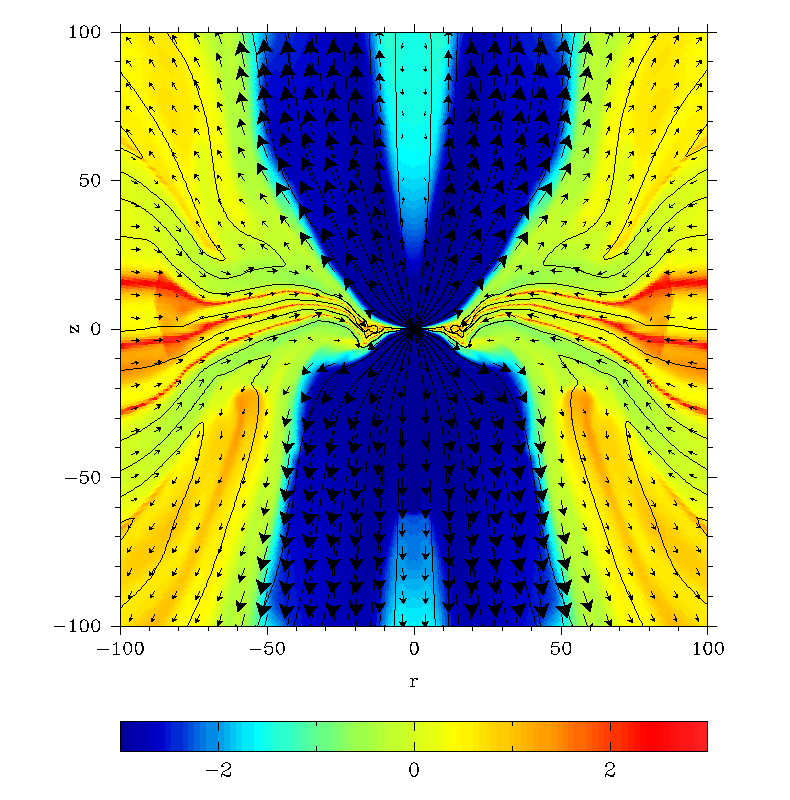}
\caption{ The magnetohydrodynamical properties of the flow: the solid
  lines show the magnetic field lines and arrows indicate the flow
  velocity (both the direction and the absolute value). The top panels
  show the distribution of the mass density ($\log \rho$ is shown by
  color) for the whole computation regions (top right panel), and for the
  zoomed central part (top left panel). The bottom panels indicate the
  ratio of the toroidal to poloidal magnetic field components ($\log
  B_{\phi}/B_{p}$ is shown by color in the bottom left panel); and the
  ratio of the gas to magnetic field pressures ($\log P_{\rm g}/P_{\rm
    m}$ is shown by color in the bottom right panel).  The initial
  magnetic field in the stellar wind was assumed to be $10^3$~G, and
  the accretion rate was selected to be $\dot{M}_{BH}=8\times10^{-11}
  M_{\odot} \mbox{yr}^{-1}$. The shown snapshot of the flow corresponds to  the
  time moment $t=0.145$~s after the accretion was initiated. }
\label{fig3}
\end{figure*}

The jets are ejected in the directions of the BH rotation axis and
blow up bubbles, as it is shown in Fig.\ref{fig3} (right top
panel). In the zoomed image of the central part, shown in
Fig.\ref{fig3} (left top panel), one can see that the matter
in-falling occurs through a narrow region in the equatorial
plane. However we note, that although this can mimic a thin accretion
disk \citep{bkr74,bkr76}, the matter velocity has no rotational
component in this region. This can be clearly seen from the left
bottom panel in Fig.\ref{fig3}, where the ratio of toroidal and
poloidal components of the B-field are shown. Any toroidal velocity
component would lead to an enhancement of the corresponding
(i.e. toroidal) magnetic field component. It is noteworthy that this
configuration with radial accretion along the equatorial plane does
not significantly differ from the standard disk accretion (which was
adopted in the conventional BZ mechanism) in the regions close to the
BH horizon since, even in the case of disk accretion, a free falling
region is formed inside the marginally stable orbit.

Another important aspect of the considered accretion regime is shown
in Fig.\ref{fig3} bottom right panel, where the ratio of gas and
magnetic pressures is shown. It can be seen that in the equatorial
region the gas pressure dominates, but the jet
itself is strongly dominated by the magnetic pressure. This
configuration fits perfectly the paradigm of the magnetically driven
jets, where the gas pressure in the jet is expected to be negligible,
while in the equatorial plane the presence of matter is required to keep a
strong magnetic field near the horizon.

\begin{figure*}
\includegraphics[width=0.48\linewidth]{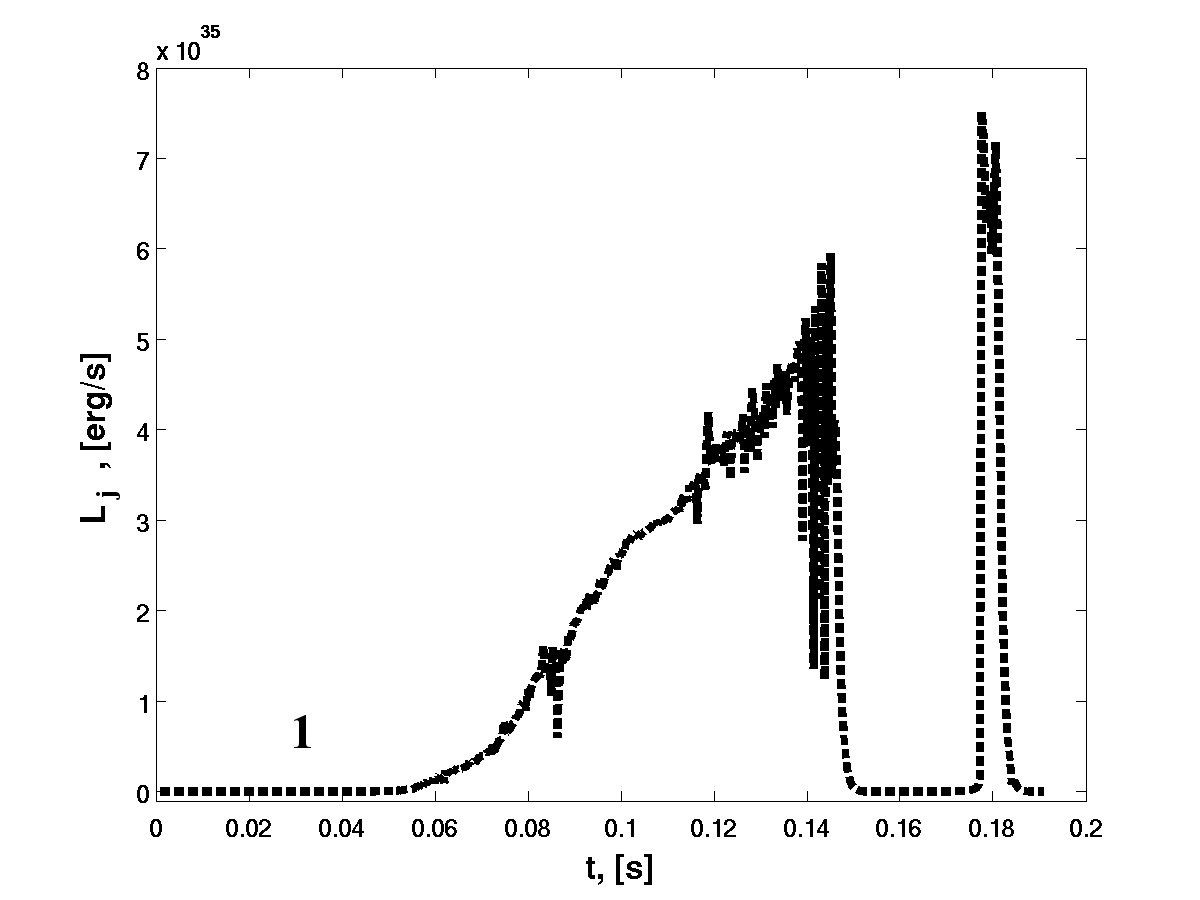}
\includegraphics[width=0.48\linewidth]{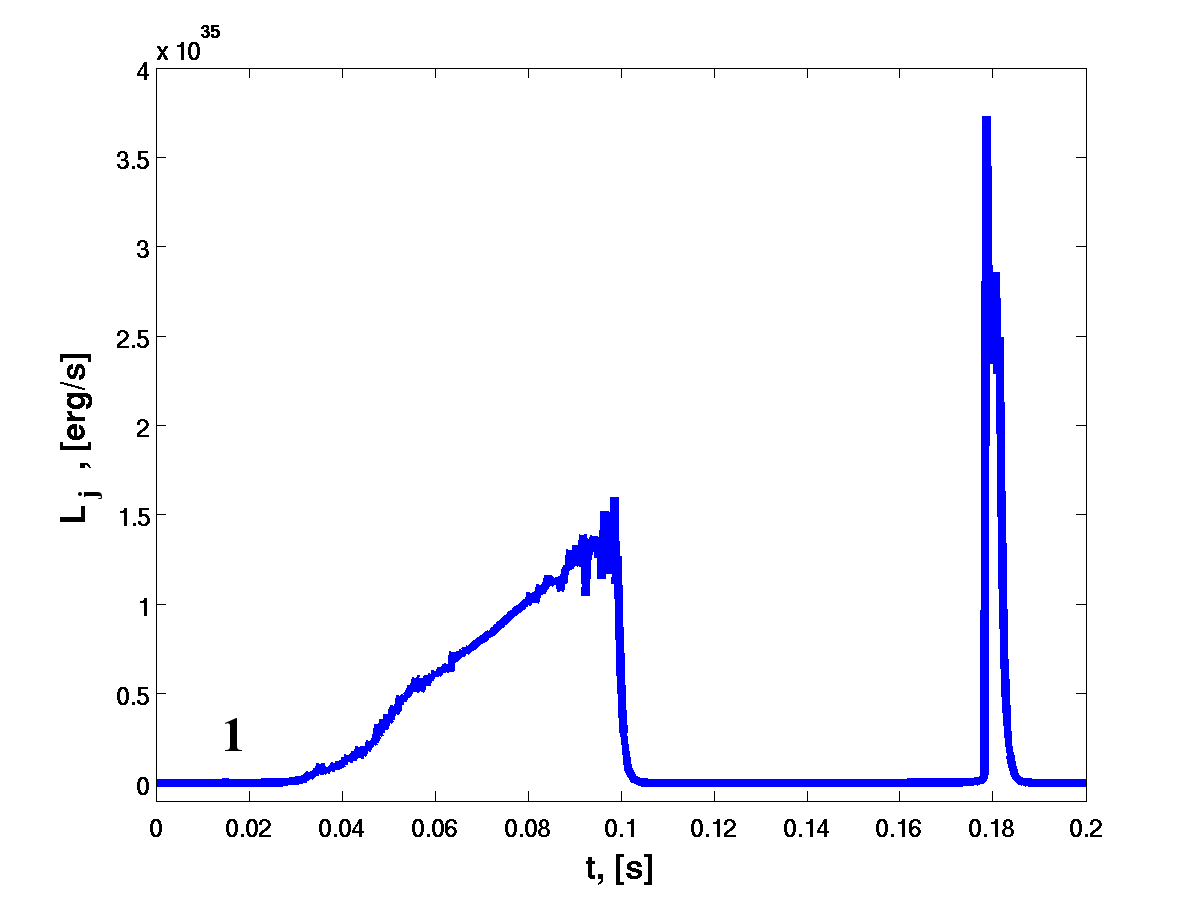}
\caption{The jet luminosity as a function of time, $t$, since the
  moment when the accretion was started. The initial magnetic field
  in the stellar wind was assumed to be $10^3$~G, and two accretion
  rates were considered $\dot{M}_{BH}=8\times10^{-11} M_{\odot}
  \mbox{yr}^{-1}$ (left panel); and $\dot{M}_{BH}=1.5\times10^{-11}
  M_{\odot} \mbox{yr}^{-1}$ (right panel).}
  \label{lbzt}
\end{figure*}

\begin{figure}
\includegraphics[width=0.98\linewidth]{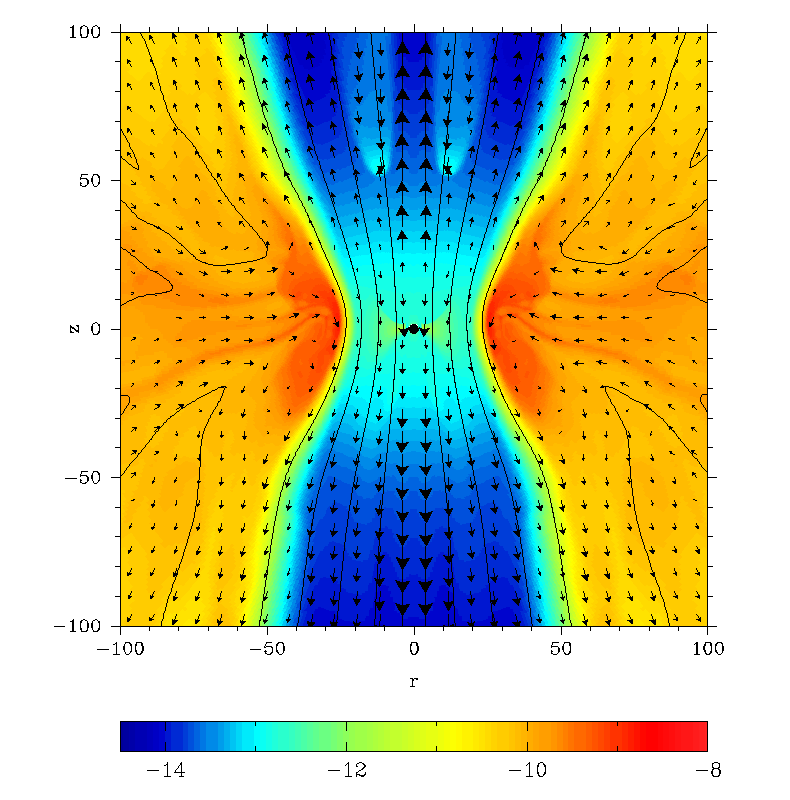}
\caption{The flow configuration corresponding to the ``magnetically
  arrested'' regime. The solid lines show the magnetic field lines and
  the arrows indicate the flow velocity (both the direction and the
  absolute value). The color corresponds to $\log \rho$. It can be
  seen that the accretion is stopped at this moment and jet ejection
  is significantly less efficient as compared to the configuration
  shown in Fig.~\ref{fig3}.  The initial magnetic field in the stellar
  wind was assumed to be $10^3$~G, and the accretion rate was selected
  to be $\dot{M}_{BH}=8\times10^{-11} M_{\odot} \mbox{yr}^{-1}$. The
  shown snapshot of the flow corresponds to the time moment
  $t=0.159$~s after the accretion was initiated (this moment
  corresponds to the ``paused jet'' as seen in Fig.~\ref{lbzt}).}
\label{fig4}
\end{figure}

The time dependence of the jet power is shown in Fig.\ref{lbzt}. It can be seen that after the
magnetic flux accumulation stage, the jet power starts to increase quickly. This growing phase is
stopped by a sudden drop of the jet power. This is caused by  
gradual increase of the magnetic field flux at the BH horizon during accretion. When the
magnetic pressure overtakes the in-falling matter ram pressure, a fast change of the structure of
the B-field occurs, leading to the formation of a {\it magnetically arrested torus}, as shown in
Fig.\ref{fig4}. A similar situation is discussed by \citet{nia03,I08} for the case of disk
accretion.

Since in the case of a {\it magnetically arrested torus} no accretion
occurs, a fast increase of the gas pressure is unavoidable. Once the
gas pressure reaches a value exceeding the B-field pressure, the
accretion, as well as the jet power supply, are resumed.  We
note however that in the case of a 3D setup a natural escape of the
magnetic flux can be realized through the generation of {\it magnetic
  tubes}. In principle, this could lead to the formation of a nearly
steady flow, although a quasi periodic pattern may still remain in the
jet power time dependence. A detailed study of this issue is beyond
the scope of this paper and will be discussed elsewhere, below we
present a qualitative discussion of this effect, which despite of 
qualitative still allows rather fundamental conclusions. 
In the frameworks of 2D calculations,
a quasi periodic consequence of ``on/off'' phases is expected. For the
computational B-field strength the typical period of these
oscillations is a fraction of a second. In fact for a large range of
the wind magnetic field value, this period depends only on the accretion
rate and BH mass (see for detail Sect.~\ref{concl}).

\section{Discussion}
\label{concl}

The obtained numerical solution may have some interesting implications for the phenomenological study
of \muq s. Namely, it was shown that an effective jet launch is possible in binary systems in the
case of the direct wind accretion, i.e. without the formation of an accretion disk. Since the
present phenomenological description of the processes occurring in \muq s relays on the
observational appearance of the accretion disk, which lacks in the case of direct wind 
accretion \muq s should fall out from this framework. Although the in-falling matter gets
strongly heated at the termination shock of the in-falling matter, given the very short free-fall and
dynamical time-scales compared to the cooling time, no detectable thermal emission is produced
in the framework of the scenario. Thus, the only way to obtain observable 
features of the suggested scenario is to study in detail the properties of the jet, and in
particular its modulation properties.


\subsection{Jet internal time-scales}

As it was discussed above the reason behind the strong modulation (``on/off'' states) of the jet
power is the accumulation of the magnetic field flux at the BH horizon. When the magnetic field
pressure exceeds the ram pressure of the accretion flow, accretion is stopped and the jet
switches off. It is possible to estimate the variability time from the comparison of the two key forces:
magnetic pressure
\begin{equation}
P_{\rm m}={B^2\over 8\pi}={\Psi^2\over 8\pi^3R^4}\,,
\label{pmeq}
\end{equation}
 and gravitational hydrostatic pressure  
\begin{equation}
\label{greq}
P_{\rm gr}=\frac\tau A{GM_{\rm BH}\dot{M}\over 4\pi R^4}\,.
\end{equation}
Here $A\le1$ is a dimensionless parameter which accounts for the flow geometry; $\tau$ is the time
since the moment when accretion is stopped,  and $G$ is the gravitational constant. Combining
these two relations one obtains the time required to restart accretion:
\begin{equation}
\label{mg_rel}
\tau={A\over 2\pi^2GM_{\rm BH}}{\Psi^2\over \dot{M}}\,.
\end{equation}
If there is no escape of the magnetic flux, i.e. the value of $\Psi$
is increasing monotonically with time $t$, the interval between jet
activations $\tau$ becomes longer with time, and asymptotically, given
the quadratic dependence on $\Psi$ in Eq.(\ref{mg_rel}), the accretion
process stops completely. Thus, the key condition for the
quasi-periodic jet is the escape of the magnetic flux from a
magnetically arrested region. We note that the escape of the magnetic
field can be studied numerically only in the a 3D approach, i.e. a
self-consistent description of this process remains out of the scope
of this paper, and should be studied elsewhere. In what follows we
adopt the critical value of $\Psi_{\rm cr}$, when the flux escapes, as
a free parameter. This parameter is in fact directly connected to the
jet power through Eq.(\ref{e-bz}), thus the escape of the magnetic
field defines not only the duration of the interval when the jet is
paused, but also its available power during the jet ejection.

In the case of non-steady jet activity, the averaged jet
luminosity is the key quantity characterizing the possible
observational appearance of such jets. Obviously, the averaged power
may depend on many different aspects of the involved processes,
e.g. on the magnetic flux escape rate. In the simplest case, the
critical magnetic flux value is smaller than the value required to
pause the accretion: $\Psi_{\rm cr}<\Psi_0$, i.e. the magnetic flux
starts to escape before the magnetic field can stop the accretion.
Then, the jet is expected to be persistent with a power of
\begin{equation}
L\sub{BZ}\approx10^{35}\Psi_{\rm cr,20}^2
\left(\frac{M_{BH}}{3 M_\odot}\right)^{-2} \,\mbox{erg}\, \mbox{s}^{-1},
\label{jet_persistent}
\end{equation}
where $\Psi_{\rm cr,20}=\Psi_{\rm cr}/10^{20}\rm G\,cm^{2}$.  Given
the assumed condition $\Psi_{\rm cr}<\Psi_0$, an upper limit for the
jet luminosity can be obtained. Our numerical simulations show that
the jet power $L_{\rm j} = C \dot{M}c^2$, where $C\sim 0.05$,
corresponds to the magnetic flux $\Psi_0$\footnote{We note that the
  value of the parameter $C$ depends on the BH rotation parameter
  $a$. Since the critical value of the magnetic flux $\Psi_{\rm cr}$
  is a weak function of $a$ \citep[see e.g.][]{tnm11}, the
  parameter $C$ should be roughly proportional to the  function $f(a)$ from
  Eq.(\ref{e-bz}). Thus, for a fast rotating BH, this parameter can have a value  up to 10 times
  larger than the one obtained in our simulations.}.  Here we note that, 
contrary to the widely accepted disk accretion paradigm, in the
considered case the accretion rate cannot be estimated from 
the thermal radiation produced by the accreted matter.

In the case if the $\Psi_{\rm cr}>\Psi_0$, one should expect a more complicated temporary patterns
in the system. Namely, the quasi-periodic behavior of the jet will not be vanished by the escape of
the magnetic field, and the jet active phases will be separated  by period of the jet silence of
duration
\begin{equation}
\label{tau_crit}
\tau_{\rm cr}={A\over 2\pi^2GM_{\rm BH}}{\Psi_{\rm cr}^2\over \dot{M}}\,.
\end{equation}
Although the interval of the jet  active phase depends on the process of the magnetic flux escape rate,
the averaged jet luminosity can be estimated assuming that the escape process is rather slow as
compared to the variability time. In this case, the duration of the active phase is roughly 
determined by time interval during which the accretion flow can keep the magnetic flux of $\Psi_{\rm cr}$
near the BH horizon.
Accounting for the phases when the accretion stops, the matter ram pressure can be estimated
as
\begin{equation}
\label{ram_pr}
P_{\rm ram}\approx{\tau_{\rm cr}+t_{\rm ac}\over t_{\rm ac}}{\dot{M}\sqrt{GM_{\rm BH}}\over 4\pi
R_{\rm G}^{5/2}}\,,
\end{equation}
where $t_{\rm ac}$ is the duration of the jet active phase, and $R_{\rm G}$ is the gravitational
radius of the BH. Using Eq.(\ref{pmeq}), one can obtain the following simple relation
\begin{equation}
\label{t_ac}
{t_{\rm ac}\over\tau_{\rm cr}+t_{\rm ac}}=\left({\Psi_0\over\Psi_{\rm cr}}\right)^2\,.
\end{equation}
Since the BZ luminosity is proportional to $\Psi^2$, the jet averaged luminosity appears to be
independent on the $\Psi_{\rm cr}$:
\begin{equation}
\label{jet_power}
<L_{\rm j}>={t_{\rm ac}\over\tau_{\rm cr}+t_{\rm ac}}\times L_{\rm BZ}\left(\Psi_{\rm
cr}\right)=L_{\rm BZ}\left(\Psi_0\right)\approx C\dot{M}c^2\,.
\end{equation}
Thus, for a broad range of parameters one can expect a universal
dependence of the jet power on the accretion rate.

\subsection{\ls as a \muq}
\label{ls}
The universal averaged jet power expected in the suggested scenario allows to make meaningful
estimates for the specific case of \ls{} without 3D modeling. Indeed, the jet power depends on the
accretion  only and for the expected rate of $\dot{M}=8\times10^{-11} M_\odot\rm yr^{-1}$ one
obtains 
\begin{equation}
\label{ls_jet}
L_{\rm j,LS}\approx 3\times10^{35}\rm \, erg\, s^{-1}\,,
\end{equation}
per each jets. This estimate is similar to flux level detected in GeV energy band with {\it
Fermi}/LAT telescope, thus to explain total non-thermal luminosity in the framework of the 
\muq scenario one needs
to invoke Doppler boosting of the emission. The Doppler boosting factor depends on the jet bulk
Lorentz factor and the orientation of the jet in respect to the observer.  The jet is orientated
along the BH rotation axis, thus most likely it is directed perpendicularly to the orbital plane.
The most feasible orbital inclination  in \ls{} is $i\sim20^\circ-25^\circ$ \citep{crr05,ssk11}, thus
one can obtain an upper limit on the boosting factor:  $\delta=1/[\Gamma(1-\beta\cos i)]\approx
2\Gamma/(1+\Gamma^2 i^2)$. If  $\Gamma> 1/i$ the emission emitted toward Earth will be de-boosted,
the optimal boosting can be achieved if $\Gamma\approx 1/i$. Thus, the largest Doppler boosting
factor can be estimated as $\delta \lesssim 3$, which should lead to the apparent jet luminosity at the
level of $<10^{37} \rm erg\,s^{-1}$. For a relatively high, $>10\%$ conversion rate of  
the jet luminosity to non-thermal emission, this energetics can explain the observed broad-band 
emission, although a dedicated  modeling is required to shed the light on the possibility 
of the production of the detected radiation in the framework of the \muq{} scenario.

\section{Summary and final remarks}
\label{sec:sum}

In this paper we study the wind accretion onto a rotating BH in close binary systems harboring a
young massive star. It is shown that the angular momentum of the accreted stellar wind material is
not sufficient for the formation of the accretion disk. On the other hand, the direct wind accretion
can create conditions for the activation of the BZ process, and a powerful jet can be launched
without any signatures of accretion, unlike it is usually seen in bright thermal X-ray binaries. The jet
power is expected to be modulated on a sub-second scale, which can lead to an efficient jet loading
with stellar wind material (see Derishev et al. 2011, in preparation).
Although the flux escape mechanism
should be properly treated with regard to the formation of persistent/quasi-periodic jets, a universal estimate on the
jet averaged power is obtained. It is expected that approximately $\sim 10\%$ of the accreted rest energy
can be chanelled into the jet. In the specific case of \ls{}, the obtained jet luminosity can be responsible 
for the observed non-thermal radiation if one invokes Doppler boosting, which can
enhanced the apparent flux from the system and to be additional factor of the modulation during the orbital motion. 
The future study of the non-thermal production in the
framework of the suggested scenario, would require a self-consistent modelings of the wind accretion
to the BH \citep{oor10}, propagation of jet through binary system environment \citep{pb08,pbk10},
and non-thermal processes occurring in the jet.

\section*{Acknowledgments}

The authors thanks to Evgeny Derishev and Valenti Bosh-Ramon for useful discussions.
BMV is thankful to State contract 2011-1.4-508-008/9 from FTP of 
RF Ministry of Education and Science.
The calculations were fulfilled at cluster of Moscow State University "Chebyshev"
and computational facilities of MPI-K Heidelberg. 



\begin{thebibliography}{}
\bibitem[\protect\citeauthoryear{{Abdo}, {Ackermann}, {Ajello}, {Allafort},
  {Ballet}, {Barbiellini}, {Bastieri} \& {et al.}}{{Abdo}
  et~al.}{2011}]{fermi_psr11a}
{Abdo} A.~A.,  {Ackermann} M.,  {Ajello} M.,  {Allafort} A.,  {Ballet} J.,
  {Barbiellini} G.,  {Bastieri} D.,    {et al.} 2011, \apjl, 736, L11+

\bibitem[\protect\citeauthoryear{{Abdo}, {Ackermann}, {Ajello}, {Atwood},
  {Axelsson}, {Baldini}, {Ballet} \& {et al.}}{{Abdo} et~al.}{2009a}]{fermi09a}
{Abdo} A.~A.,  {Ackermann} M.,  {Ajello} M.,  {Atwood} W.~B.,  {Axelsson} M.,
  {Baldini} L.,  {Ballet} J.,    {et al.} 2009a, \apjl, 701, L123

\bibitem[\protect\citeauthoryear{{Abdo}, {Ackermann}, {Ajello}, {Atwood},
  {Axelsson}, {Baldini}, {Ballet} \& {et al.}}{{Abdo} et~al.}{2009b}]{fermi09}
{Abdo} A.~A.,  {Ackermann} M.,  {Ajello} M.,  {Atwood} W.~B.,  {Axelsson} M.,
  {Baldini} L.,  {Ballet} J.,    {et al.} 2009b, \apjl, 706, L56

\bibitem[\protect\citeauthoryear{{Acciari}, {Aliu}, {Arlen}, {Aune},
  {Beilicke}, {Benbow}, {Bradbury} \& {et al.}}{{Acciari}
  et~al.}{2011}]{veritas11}
{Acciari} V.~A.,  {Aliu} E.,  {Arlen} T.,  {Aune} T.,  {Beilicke} M.,  {Benbow}
  W.,  {Bradbury} S.~M.,    {et al.} 2011, \apj, 738, 3

\bibitem[\protect\citeauthoryear{{Acciari}, {Aliu}, {Arlen}, {Bautista},
  {Beilicke}, {Benbow}, {B{\"o}ttcher} \& {et al.}}{{Acciari}
  et~al.}{2009}]{veritas09}
{Acciari} V.~A.,  {Aliu} E.,  {Arlen} T.,  {Bautista} M.,  {Beilicke} M.,
  {Benbow} W.,  {B{\"o}ttcher} M.,    {et al.} 2009, \apj, 700, 1034

\bibitem[\protect\citeauthoryear{{Acciari}, {Beilicke}, {Blaylock}, {Bradbury},
  {Buckley}, {Bugaev}, {Butt} \& {et al.}}{{Acciari} et~al.}{2008}]{veritas08}
{Acciari} V.~A.,  {Beilicke} M.,  {Blaylock} G.,  {Bradbury} S.~M.,  {Buckley}
  J.~H.,  {Bugaev} V.,  {Butt} Y.,    {et al.} 2008, \apj, 679, 1427

\bibitem[\protect\citeauthoryear{{Aharonian}, {Akhperjanian}, {Anton}, {Barres
  de Almeida}, {Bazer-Bachi}, {Becherini}, {Behera} \& {et al.}}{{Aharonian}
  et~al.}{2009}]{hess09}
{Aharonian} F.,  {Akhperjanian} A.~G.,  {Anton} G.,  {Barres de Almeida} U.,
  {Bazer-Bachi} A.~R.,  {Becherini} Y.,  {Behera} B.,    {et al.} 2009, \aap,
  507, 389

\bibitem[\protect\citeauthoryear{{Aharonian}, {Akhperjanian}, {Aye},
  {Bazer-Bachi}, {Beilicke}, {Benbow}, {Berge} \& {et al.}}{{Aharonian}
  et~al.}{2005a}]{hess05a}
{Aharonian} F.,  {Akhperjanian} A.~G.,  {Aye} K.-M.,  {Bazer-Bachi} A.~R.,
  {Beilicke} M.,  {Benbow} W.,  {Berge} D.,    {et al.} 2005a, \aap, 442, 1

\bibitem[\protect\citeauthoryear{{Aharonian}, {Akhperjanian}, {Aye},
  {Bazer-Bachi}, {Beilicke}, {Benbow}, {Berge} \& {et al.}}{{Aharonian}
  et~al.}{2005b}]{hess05}
{Aharonian} F.,  {Akhperjanian} A.~G.,  {Aye} K.-M.,  {Bazer-Bachi} A.~R.,
  {Beilicke} M.,  {Benbow} W.,  {Berge} D.,    {et al.} 2005b, Science, 309,
  746

\bibitem[\protect\citeauthoryear{{Aharonian}, {Akhperjanian}, {Bazer-Bachi},
  {Beilicke}, {Benbow}, {Berge}, {Bernl{\"o}hr}, {Boisson} \& {et
  al.}}{{Aharonian} et~al.}{2006}]{hess06}
{Aharonian} F.,  {Akhperjanian} A.~G.,  {Bazer-Bachi} A.~R.,  {Beilicke} M.,
  {Benbow} W.,  {Berge} D.,  {Bernl{\"o}hr} K.,  {Boisson} C.,    {et al.}
  2006, \aap, 460, 743

\bibitem[\protect\citeauthoryear{{Albert}, {Aliu}, {Anderhub}, {Antonelli},
  {Antoranz}, {Backes}, {Baixeras} \& {et al.}}{{Albert}
  et~al.}{2009}]{magic09}
{Albert} J.,  {Aliu} E.,  {Anderhub} H.,  {Antonelli} L.~A.,  {Antoranz} P.,
  {Backes} M.,  {Baixeras} C.,    {et al.} 2009, \apj, 693, 303

\bibitem[\protect\citeauthoryear{{Albert}, {Aliu}, {Anderhub}, {Antoranz},
  {Armada}, {Asensio}, {Baixeras} \& {et al.}}{{Albert} et~al.}{2006}]{magic06}
{Albert} J.,  {Aliu} E.,  {Anderhub} H.,  {Antoranz} P.,  {Armada} A.,
  {Asensio} M.,  {Baixeras} C.,    {et al.} 2006, Science, 312, 1771

\bibitem[\protect\citeauthoryear{{Albert}, {Aliu}, {Anderhub}, {Antoranz},
  {Armada}, {Baixeras}, {Barrio} \& {et al.}}{{Albert} et~al.}{2007}]{magic07}
{Albert} J.,  {Aliu} E.,  {Anderhub} H.,  {Antoranz} P.,  {Armada} A.,
  {Baixeras} C.,  {Barrio} J.~A.,    {et al.} 2007, \apjl, 665, L51

\bibitem[\protect\citeauthoryear{{Artemova}, {Bisnovatyi-Kogan}, {Igumenshchev}
  \& {Novikov}}{{Artemova} et~al.}{2006}]{abin06}
{Artemova} Y.~V.,  {Bisnovatyi-Kogan} G.~S.,  {Igumenshchev} I.~V.,
  {Novikov} I.~D.,  2006, \apj, 637, 968

\bibitem[\protect\citeauthoryear{{Bardeen}, {Press} \& {Teukolsky}}{{Bardeen}
  et~al.}{1972}]{bpt72}
{Bardeen} J.~M.,  {Press} W.~H.,    {Teukolsky} S.~A.,  1972, \apj, 178, 347

\bibitem[\protect\citeauthoryear{{Barkov} \& {Komissarov}}{{Barkov} \&
  {Komissarov}}{2008a}]{BK08c}
{Barkov} M.~V.,  {Komissarov} S.~S.,  2008a, in {F.~A.~Aharonian, W.~Hofmann,
  \& F.~Rieger} ed., American Institute of Physics Conference Series Vol.~1085
  of American Institute of Physics Conference Series, {Central engines of Gamma
  Ray Bursts. Magnetic mechanism in the collapsar model.}.
pp 608--611

\bibitem[\protect\citeauthoryear{{Barkov} \& {Komissarov}}{{Barkov} \&
  {Komissarov}}{2008b}]{BK08b}
{Barkov} M.~V.,  {Komissarov} S.~S.,  2008b, \mnras, 385, L28

\bibitem[\protect\citeauthoryear{{Barkov} \& {Komissarov}}{{Barkov} \&
  {Komissarov}}{2010}]{BK10}
{Barkov} M.~V.,  {Komissarov} S.~S.,  2010, \mnras, 401, 1644

\bibitem[\protect\citeauthoryear{{Barkov} \& {Komissarov}}{{Barkov} \&
  {Komissarov}}{2011}]{BK11rns}
{Barkov} M.~V.,  {Komissarov} S.~S.,  2011, \mnras, 415, 944

\bibitem[\protect\citeauthoryear{{Bisnovatyi-Kogan} \&
  {Ruzmaikin}}{{Bisnovatyi-Kogan} \& {Ruzmaikin}}{1974}]{bkr74}
{Bisnovatyi-Kogan} G.~S.,  {Ruzmaikin} A.~A.,  1974, \apss, 28, 45

\bibitem[\protect\citeauthoryear{{Bisnovatyi-Kogan} \&
  {Ruzmaikin}}{{Bisnovatyi-Kogan} \& {Ruzmaikin}}{1976}]{bkr76}
{Bisnovatyi-Kogan} G.~S.,  {Ruzmaikin} A.~A.,  1976, \apss, 42, 401

\bibitem[\protect\citeauthoryear{{Blandford} \& {Begelman}}{{Blandford} \&
  {Begelman}}{1999}]{bb99}
{Blandford} R.~D.,  {Begelman} M.~C.,  1999, \mnras, 303, L1

\bibitem[\protect\citeauthoryear{{Blandford} \& {Znajek}}{{Blandford} \&
  {Znajek}}{1977}]{BZ77}
{Blandford} R.~D.,  {Znajek} R.~L.,  1977, \mnras, 179, 433

\bibitem[\protect\citeauthoryear{{Bogovalov}, {Khangulyan}, {Koldoba},
  {Ustyugova} \& {Aharonian}}{{Bogovalov} et~al.}{2011}]{bkk11}
{Bogovalov} S.,  {Khangulyan} D.,  {Koldoba} A.~V.,  {Ustyugova} G.~V.,
  {Aharonian} F.~A.,  2011, ArXiv e-prints

\bibitem[\protect\citeauthoryear{{Bogovalov}, {Khangulyan}, {Koldoba},
  {Ustyugova} \& {Aharonian}}{{Bogovalov} et~al.}{2008}]{bkk08}
{Bogovalov} S.~V.,  {Khangulyan} D.~V.,  {Koldoba} A.~V.,  {Ustyugova} G.~V.,
   {Aharonian} F.~A.,  2008, \mnras, 387, 63

\bibitem[\protect\citeauthoryear{{Bosch-Ramon} \& {Barkov}}{{Bosch-Ramon} \&
  {Barkov}}{2011}]{brb11}
{Bosch-Ramon} V.,  {Barkov} M.~V.,  2011, ArXiv e-prints

\bibitem[\protect\citeauthoryear{{Bosch-Ramon} \& {Khangulyan}}{{Bosch-Ramon}
  \& {Khangulyan}}{2009}]{bk09}
{Bosch-Ramon} V.,  {Khangulyan} D.,  2009, International Journal of Modern
  Physics D, 18, 347

\bibitem[\protect\citeauthoryear{{Bosch-Ramon} \& {Khangulyan}}{{Bosch-Ramon}
  \& {Khangulyan}}{2011}]{bk11}
{Bosch-Ramon} V.,  {Khangulyan} D.,  2011, ArXiv e-prints

\bibitem[\protect\citeauthoryear{{Bosch-Ramon}, {Khangulyan} \&
  {Aharonian}}{{Bosch-Ramon} et~al.}{2008}]{bka08}
{Bosch-Ramon} V.,  {Khangulyan} D.,    {Aharonian} F.~A.,  2008, \aap, 482, 397

\bibitem[\protect\citeauthoryear{{Casares}}{{Casares}}{2007}]{cas07}
{Casares} J.,  2007, in {V.~Karas \& G.~Matt} ed., IAU Symposium Vol.~238 of
  IAU Symposium, {Observational evidence for stellar-mass black holes}.
pp 3--12

\bibitem[\protect\citeauthoryear{{Casares}, {Rib{\'o}}, {Ribas}, {Paredes},
  {Mart{\'i}} \& {Herrero}}{{Casares} et~al.}{2005}]{crr05}
{Casares} J.,  {Rib{\'o}} M.,  {Ribas} I.,  {Paredes} J.~M.,  {Mart{\'i}} J.,
   {Herrero} A.,  2005, \mnras, 364, 899

\bibitem[\protect\citeauthoryear{{Davies} \& {Pringle}}{{Davies} \&
  {Pringle}}{1980}]{dp80}
{Davies} R.~E.,  {Pringle} J.~E.,  1980, \mnras, 191, 599


\bibitem[\protect\citeauthoryear{{Dhawan}, {Mioduszewski} \& {Rupen}}{{Dhawan}
  et~al.}{2006}]{dmr06}
{Dhawan} V.,  {Mioduszewski} A.,    {Rupen} M.,  2006, in VI Microquasar
  Workshop: Microquasars and Beyond {LS I +61 303 is a Be-Pulsar binary, not a
  Microquasar}

\bibitem[\protect\citeauthoryear{{Dubus}}{{Dubus}}{2006}]{dub06}
{Dubus} G.,  2006, \aap, 456, 801

\bibitem[\protect\citeauthoryear{{Dubus}, {Cerutti} \& {Henri}}{{Dubus}
  et~al.}{2010}]{dch10}
{Dubus} G.,  {Cerutti} B.,    {Henri} G.,  2010, \aap, 516, A18+

\bibitem[\protect\citeauthoryear{{Esin}, {Narayan}, {Cui}, {Grove} \&
  {Zhang}}{{Esin} et~al.}{1998}]{enc98}
{Esin} A.~A.,  {Narayan} R.,  {Cui} W.,  {Grove} J.~E.,    {Zhang} S.-N.,
  1998, \apj, 505, 854

\bibitem[\protect\citeauthoryear{{Igumenshchev}}{{Igumenshchev}}{2008}]{I08}
{Igumenshchev} I.~V.,  2008, \apj, 677, 317

\bibitem[\protect\citeauthoryear{{Illarionov} \& {Sunyaev}}{{Illarionov} \&
  {Sunyaev}}{1975}]{is75}
{Illarionov} A.~F.,  {Sunyaev} R.~A.,  1975, {\aa}p, 39, 185

\bibitem[\protect\citeauthoryear{{Ishii}, {Matsuda}, {Shima}, {Livio}, {Anzer}
  \& {Boerner}}{{Ishii} et~al.}{1993}]{im93}
{Ishii} T.,  {Matsuda} T.,  {Shima} E.,  {Livio} M.,  {Anzer} U.,    {Boerner}
  G.,  1993, \apj, 404, 706

\bibitem[\protect\citeauthoryear{{Johnston}, {Manchester}, {Lyne}, {D'Amico},
  {Bailes}, {Gaensler} \& {Nicastro}}{{Johnston} et~al.}{1996}]{jml96}
{Johnston} S.,  {Manchester} R.~N.,  {Lyne} A.~G.,  {D'Amico} N.,  {Bailes} M.,
   {Gaensler} B.~M.,    {Nicastro} L.,  1996, \mnras, 279, 1026

\bibitem[\protect\citeauthoryear{{Kaisig}, {Tajima} \& {Lovelace}}{{Kaisig}
  et~al.}{1992}]{ktl92}
{Kaisig} M.,  {Tajima} T.,    {Lovelace} R.~V.~E.,  1992, \apj, 386, 83

\bibitem[\protect\citeauthoryear{{Kennel} \& {Coroniti}}{{Kennel} \&
  {Coroniti}}{1984}]{kc84a}
{Kennel} C.~F.,  {Coroniti} F.~V.,  1984, \apj, 283, 710

\bibitem[\protect\citeauthoryear{{Khangulyan}, {Aharonian}, {Bogovalov} \&
  {Ribo}}{{Khangulyan} et~al.}{2011}]{kab11a}
{Khangulyan} D.,  {Aharonian} F.~A.,  {Bogovalov} S.~V.,    {Ribo} M.,  2011,
  ArXiv e-prints

\bibitem[\protect\citeauthoryear{{Khangulyan}, {Hnatic}, {Aharonian} \&
  {Bogovalov}}{{Khangulyan} et~al.}{2007}]{kha07}
{Khangulyan} D.,  {Hnatic} S.,  {Aharonian} F.,    {Bogovalov} S.,  2007,
  \mnras, 380, 320

\bibitem[\protect\citeauthoryear{{Khangulyan}, {Aharonian}, {Bogovalov},
  {Koldoba} \& {Ustyugova}}{{Khangulyan} et~al.}{2008}]{kab08}
{Khangulyan} D.~V.,  {Aharonian} F.~A.,  {Bogovalov} S.~V.,  {Koldoba} A.~V.,
   {Ustyugova} G.~V.,  2008, International Journal of Modern Physics D, 17,
  1909

\bibitem[\protect\citeauthoryear{{Kirk}, {Ball} \& {Skjaeraasen}}{{Kirk}
  et~al.}{2000}]{kbs00}
{Kirk} J.~G.,  {Ball} L.,    {Skjaeraasen} O.,  2000, in {M.~Kramer, N.~Wex, \&
  R.~Wielebinski} ed., IAU Colloq. 177: Pulsar Astronomy - 2000 and Beyond
  Vol.~202 of Astronomical Society of the Pacific Conference Series,
  {Predictions of inverse Compton radiation from PSR B1259-63}.
pp 531--+

\bibitem[\protect\citeauthoryear{{Kishishita}, {Tanaka}, {Uchiyama} \&
  {Takahashi}}{{Kishishita} et~al.}{2009}]{ktu09}
{Kishishita} T.,  {Tanaka} T.,  {Uchiyama} Y.,    {Takahashi} T.,  2009, \apjl,
  697, L1

\bibitem[\protect\citeauthoryear{{Komissarov}}{{Komissarov}}{1999}]{K99}
{Komissarov} S.~S.,  1999, \mnras, 303, 343

\bibitem[\protect\citeauthoryear{{Komissarov}}{{Komissarov}}{2004}]{K04b}
{Komissarov} S.~S.,  2004, \mnras, 350, 427

\bibitem[\protect\citeauthoryear{{Komissarov}}{{Komissarov}}{2006}]{K06}
{Komissarov} S.~S.,  2006, \mnras, 368, 993

\bibitem[\protect\citeauthoryear{{Komissarov} \& {Barkov}}{{Komissarov} \&
  {Barkov}}{2007}]{KB07}
{Komissarov} S.~S.,  {Barkov} M.~V.,  2007, \mnras, 382, 1029

\bibitem[\protect\citeauthoryear{{Komissarov} \& {Barkov}}{{Komissarov} \&
  {Barkov}}{2009}]{kb09}
{Komissarov} S.~S.,  {Barkov} M.~V.,  2009, \mnras, 397, 1153

\bibitem[\protect\citeauthoryear{{Kudritzki} \& {Puls}}{{Kudritzki} \&
  {Puls}}{2000}]{kp00sw}
{Kudritzki} R.,  {Puls} J.,  2000, \araa, 38, 613

\bibitem[\protect\citeauthoryear{{Li}, {Torres}, {Zhang}, {Chen}, {Hadasch},
  {Ray}, {Kretschmar}, {Rea} \& {Wang}}{{Li} et~al.}{2011}]{ltz11}
{Li} J.,  {Torres} D.~F.,  {Zhang} S.,  {Chen} Y.,  {Hadasch} D.,  {Ray} P.~S.,
   {Kretschmar} P.,  {Rea} N.,    {Wang} J.,  2011, \apj, 733, 89

\bibitem[\protect\citeauthoryear{{Lovelace}}{{Lovelace}}{1976}]{lov76}
{Lovelace} R.~V.~E.,  1976, \nat, 262, 649

\bibitem[\protect\citeauthoryear{{Mold{\'o}n}, {Johnston}, {Rib{\'o}},
  {Paredes} \& {Deller}}{{Mold{\'o}n} et~al.}{2011a}]{mjr11}
{Mold{\'o}n} J.,  {Johnston} S.,  {Rib{\'o}} M.,  {Paredes} J.~M.,    {Deller}
  A.~T.,  2011a, \apjl, 732, L10+

\bibitem[\protect\citeauthoryear{{Mold{\'o}n}, {Rib{\'o}} \&
  {Paredes}}{{Mold{\'o}n} et~al.}{2011b}]{mrp11}
{Mold{\'o}n} J.,  {Rib{\'o}} M.,    {Paredes} J.~M.,  2011b, \aap, 533, L7+

\bibitem[\protect\citeauthoryear{{Narayan}, {Igumenshchev} \&
  {Abramowicz}}{{Narayan} et~al.}{2003}]{nia03}
{Narayan} R.,  {Igumenshchev} I.~V.,    {Abramowicz} M.~A.,  2003, \pasj, 55,
  L69

\bibitem[\protect\citeauthoryear{{Narayan} \& {Yi}}{{Narayan} \&
  {Yi}}{1994}]{ny94}
{Narayan} R.,  {Yi} I.,  1994, \apjl, 428, L13

\bibitem[\protect\citeauthoryear{{Negueruela}, {Rib{\'o}}, {Herrero},
  {Lorenzo}, {Khangulyan} \& {Aharonian}}{{Negueruela} et~al.}{2011}]{nrh11}
{Negueruela} I.,  {Rib{\'o}} M.,  {Herrero} A.,  {Lorenzo} J.,  {Khangulyan}
  D.,    {Aharonian} F.~A.,  2011, \apjl, 732, L11+

\bibitem[\protect\citeauthoryear{{Owocki}, {Okazaki} \& {Romero}}{{Owocki}
  et~al.}{2011}]{oor10}
{Owocki} S.~P.,  {Okazaki} A.~T.,    {Romero} G.,  2011, in {C.~Neiner,
  G.~Wade, G.~Meynet, \& G.~Peters} ed., IAU Symposium Vol.~272 of IAU
  Symposium, {Modeling TeV {$\gamma$}-rays from LS 5039: an active OB star at
  the extreme}.
pp 587--592

\bibitem[\protect\citeauthoryear{{Perucho} \& {Bosch-Ramon}}{{Perucho} \&
  {Bosch-Ramon}}{2008}]{pb08}
{Perucho} M.,  {Bosch-Ramon} V.,  2008, \aap, 482, 917

\bibitem[\protect\citeauthoryear{{Perucho}, {Bosch-Ramon} \&
  {Khangulyan}}{{Perucho} et~al.}{2010}]{pbk10}
{Perucho} M.,  {Bosch-Ramon} V.,    {Khangulyan} D.,  2010, \aap, 512, L4+

\bibitem[\protect\citeauthoryear{{Rea}, {Torres}, {Caliandro}, {Hadasch}, {van
  der Klis}, {Jonker}, {M{\'e}ndez} \& {Sierpowska-Bartosik}}{{Rea}
  et~al.}{2011}]{rtc11}
{Rea} N.,  {Torres} D.~F.,  {Caliandro} G.~A.,  {Hadasch} D.,  {van der Klis}
  M.,  {Jonker} P.~G.,  {M{\'e}ndez} M.,    {Sierpowska-Bartosik} A.,  2011,
  \mnras, 416, 1514

\bibitem[\protect\citeauthoryear{{Rea}, {Torres}, {van der Klis}, {Jonker},
  {M{\'e}ndez} \& {Sierpowska-Bartosik}}{{Rea} et~al.}{2010}]{rtk10}
{Rea} N.,  {Torres} D.~F.,  {van der Klis} M.,  {Jonker} P.~G.,  {M{\'e}ndez}
  M.,    {Sierpowska-Bartosik} A.,  2010, \mnras, 405, 2206

\bibitem[\protect\citeauthoryear{{Rib{\'o}}, {Paredes}, {Mold{\'o}n},
  {Mart{\'{\i}}} \& {Massi}}{{Rib{\'o}} et~al.}{2008}]{rpm08}
{Rib{\'o}} M.,  {Paredes} J.~M.,  {Mold{\'o}n} J.,  {Mart{\'{\i}}} J.,
  {Massi} M.,  2008, \aap, 481, 17

\bibitem[\protect\citeauthoryear{{Romanova}, {Ustyugova}, {Koldoba} \&
  {Lovelace}}{{Romanova} et~al.}{2009}]{ruk09}
{Romanova} M.~M.,  {Ustyugova} G.~V.,  {Koldoba} A.~V.,    {Lovelace} R.~V.~E.,
   2009, \mnras, 399, 1802

\bibitem[\protect\citeauthoryear{{Ruffert}}{{Ruffert}}{1997}]{r97}
{Ruffert} M.,  1997, {\aa}p, 317, 793

\bibitem[\protect\citeauthoryear{{Ruffert}}{{Ruffert}}{1999}]{r99}
{Ruffert} M.,  1999, {\aa}p, 346, 861

\bibitem[\protect\citeauthoryear{Ruffini \& Wilson}{Ruffini \&
  Wilson}{1975}]{rw75}
Ruffini R.,  Wilson J.~R.,  1975, Phys. Rev. D, 12, 2959

\bibitem[\protect\citeauthoryear{{Sarty}, {Szalai}, {Kiss}, {Matthews}, {Wu},
  {Kuschnig}, {Guenther}, {Moffat}, {Rucinski}, {Sasselov}, {Weiss}, {Huziak},
  {Johnston}, {Phillips} \& {Ashley}}{{Sarty} et~al.}{2011}]{ssk11}
{Sarty} G.~E.,  {Szalai} T.,  {Kiss} L.~L.,  {Matthews} J.~M.,  {Wu} K.,
  {Kuschnig} R.,  {Guenther} D.~B.,  {Moffat} A.~F.~J.,  {Rucinski} S.~M.,
  {Sasselov} D.,  {Weiss} W.~W.,  {Huziak} R.,  {Johnston} H.~M.,  {Phillips}
  A.,    {Ashley} M.~C.~B.,  2011, \mnras, 411, 1293

\bibitem[\protect\citeauthoryear{{Takahashi}, {Kishishita}, {Uchiyama},
  {Tanaka}, {Yamaoka}, {Khangulyan}, {Aharonian}, {Bosch-Ramon} \&
  {Hinton}}{{Takahashi} et~al.}{2009}]{tku09}
{Takahashi} T.,  {Kishishita} T.,  {Uchiyama} Y.,  {Tanaka} T.,  {Yamaoka} K.,
  {Khangulyan} D.,  {Aharonian} F.~A.,  {Bosch-Ramon} V.,    {Hinton} J.~A.,
  2009, \apj, 697, 592

\bibitem[\protect\citeauthoryear{{Tam}, {Huang}, {Takata}, {Hui}, {Kong} \&
  {Cheng}}{{Tam} et~al.}{2011}]{fermi_psr11}
{Tam} P.~H.~T.,  {Huang} R.~H.~H.,  {Takata} J.,  {Hui} C.~Y.,  {Kong}
  A.~K.~H.,    {Cheng} K.~S.,  2011, \apjl, 736, L10+

\bibitem[\protect\citeauthoryear{{Tchekhovskoy}, {Narayan} \&
  {McKinney}}{{Tchekhovskoy} et~al.}{2011}]{tnm11}
{Tchekhovskoy} A.,  {Narayan} R.,    {McKinney} J.~C.,  2011, ArXiv e-prints

\bibitem[\protect\citeauthoryear{{Uchiyama}, {Tanaka}, {Takahashi}, {Mori} \&
  {Nakazawa}}{{Uchiyama} et~al.}{2009}]{utt09}
{Uchiyama} Y.,  {Tanaka} T.,  {Takahashi} T.,  {Mori} K.,    {Nakazawa} K.,
  2009, \apj, 698, 911

\bibitem[\protect\citeauthoryear{{Usov} \& {Melrose}}{{Usov} \&
  {Melrose}}{1992}]{um92}
{Usov} V.~V.,  {Melrose} D.~B.,  1992, \apj, 395, 575

\bibitem[\protect\citeauthoryear{{Zabalza}, {Bosch-Ramon} \&
  {Paredes}}{{Zabalza} et~al.}{2011}]{zbp11}
{Zabalza} V.,  {Bosch-Ramon} V.,    {Paredes} J.~M.,  2011, ArXiv e-prints

\bibitem[\protect\citeauthoryear{{Zi{\'o}{\l}kowski}}{{Zi{\'o}{\l}kowski}}{200%
5}]{zio05}
{Zi{\'o}{\l}kowski} J.,  2005, \mnras, 358, 851

\end{thebibliography}

\end{document}